# Shock, Communication, and Yield Curve Repricing: A Two-Step Empirical Framework for Copom Events in Brazil

*Gabriel de Macedo Santos*

*Instituto de Tecnologia e Liderança*

**Abstract.** This paper proposes a two-step empirical framework to study the repricing of the Brazilian DI curve around Copom-related events. The empirical strategy separates the initial market reaction associated with the underlying shock from the subsequent repricing observed between the shock and the first Copom statement that follows it. The dataset combines a hand-built event calendar, daily market data, Focus expectations, and structured textual features extracted from Copom statements, including tone, forward-guidance direction and explicitness, and uncertainty indicators. In the updated sample, 59 events retain both analytical windows, allowing the second stage to include the full set of same-day Copom events. Baseline results suggest that the framework is most informative at the front and intermediate sections of the curve, especially for the DI 252d maturity, for which the baseline OLS specification reaches an in-sample $R^2$ of about 0.43. By contrast, explanatory power is materially weaker for the DI 504d maturity and for slope adjustments, and out-of-sample performance remains limited. The textual variables display economically plausible signs, but their statistical contribution is not uniformly robust across specifications. The main contribution of the paper is therefore methodological and applied: it offers an implementable event-based decomposition for assessing how shocks and Copom communication jointly shape curve dynamics in Brazil.

## 1. Introduction

Understanding how the Brazilian yield curve reacts around Copom-related events requires more than observing the announcement day in isolation. In practice, market pricing often starts before the statement is released, because investors first absorb the underlying shock that motivates a repricing episode and only then process the central bank communication that follows. A daily event-study design that ignores this sequence risks mixing distinct mechanisms: the immediate reaction to the shock itself and the subsequent adjustment associated with the content of the Copom statement.

This paper proposes a two-step empirical framework to address that problem. In the first step, we measure the initial repricing observed between the last market date before the shock and the first market date on or after the shock. In the second step, we measure the subsequent repricing between that effective shock date and the first market date after the Copom statement that follows the event. This design is particularly relevant in Brazil because Copom decisions are released on the same day as the decision and, under current institutional rules, the statement is published only after the second session of the meeting, from 18:30 onward. That timing implies that same-day Copom communications should affect the next trading day rather than the same-day close.

The paper is not presented as a sharp causal identification exercise. The second-stage window is still a daily communication window, not an intraday monetary surprise measure, and it can therefore absorb information that arrives between the shock and the statement. The contribution is instead an applied decomposition framework: it organizes the event around two interpretable stages and asks whether structured communication variables help explain the repricing that remains after the initial shock has already moved the curve.

This positioning connects the paper to three literatures. First, it relates to the broad literature on central bank communication, which documents that policy statements and communication design matter for market expectations. Second, it connects to event-study papers that distinguish policy decisions from statements and forward-guidance effects. Third, it relates to the growing literature that turns central bank text into structured economic variables rather than treating language as an opaque statistical object.

Using the updated project outputs, the complete event-level sample contains 59 observations from 2016-08-31 to 2026-03-18. The improvement relative to the earlier version is substantive: both analytical windows are now available for all 59 retained events, including 16 events classified as monetary policy/Copom. This correction matters because a framework meant to study Copom communication should preserve, rather than lose, same-day Copom events in the communication stage.

The main empirical findings are cautious but informative. The baseline OLS-HC3 specification performs best for the DI 252d maturity, with an in-sample $R^2$ of 0.432 and sign accuracy close to 75%. The corresponding results for the DI 504d maturity and for the 21d–504d slope are much weaker. The textual variables display economically intuitive signs in several specifications, especially for tone and guidance in the DI 252d equation, but they are not uniformly statistically significant. In addition, cross-validated performance remains modest, which reinforces the interpretation of the paper as a descriptive and applied decomposition framework rather than a forecasting model.

*Figure 1: schematic timeline of the two-step event framework.*

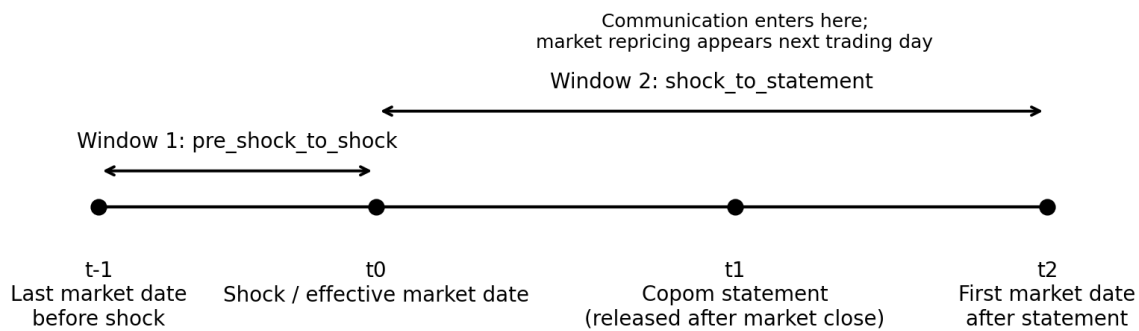

## 2. Related Literature

The first literature strand studies central bank communication more broadly. A useful starting point is Blinder, Ehrmann, Fratzscher, De Haan, and Jansen (2008), who survey the theory and evidence and show that communication has become a central part of the monetary policy toolkit. Their survey is especially relevant for the present paper because it frames communication research as primarily empirical and emphasizes that statements can shape expectations even when the policy instrument itself is unchanged.

A second strand separates policy actions from policy communication in event-study settings. Gürkaynak, Sack, and Swanson (2005) show that U.S. monetary policy announcements cannot be summarized by a single surprise factor and that an additional factor linked to the future policy path is closely associated with statements. For the euro area, Brand, Buncic, and Turunen (2010) use intraday data to distinguish decision news from communication news and find that communication exerts a sizeable effect on medium- and long-term yields. The present paper does not replicate that intraday identification strategy. Instead, it adapts the same economic intuition to a daily Brazilian setting in which the objective is an implementable decomposition rather than a clean structural identification of communication shocks.

A third strand converts policy text into economic variables. Hansen and McMahon (2016) show that computational linguistic tools can be used to measure the information that central banks release about current conditions and future policy guidance. Hubert and Labondance (2021) further show that the tone of monetary policy statements contains information for financial market participants beyond standard announcement-day variables. These studies support the choice made here to focus on structured and interpretable textual features—tone, guidance, and uncertainty—rather than on an opaque general-purpose language score.

The Brazilian literature provides the closest benchmark. Costa Filho and Rocha (2010) show that Central Bank communication affects interest rates and volatility in Brazil. Cabral and Guimarães (2015) classify the semantic content of Copom statements and show that the statements contain information about future policy and asset prices. Chague, De-Losso, Giovannetti, and Manoel (2015) find that the content of Copom minutes affects the term structure of interest rates and the volatility of future interest rates. More recently, Cavaca and Meurer (2024) show that asymmetry and uncertainty matter for the response of the Brazilian yield curve to monetary policy, while Alves, Abraham, and Laurini (2023) and Alves and Laurini (2025) show that communication variables can help explain or predict movements in the Brazilian term structure.

Relative to that literature, the present paper makes a narrower but operational contribution. It does not attempt to produce a definitive measure of monetary policy surprises, nor does it claim a forecasting breakthrough. Its main contribution is to organize event information into a two-step daily framework that is transparent, reproducible, and directly usable in macro-financial research and market analysis.

## 3. Data and Event Construction

The project starts from 75 raw events. After restricting the analysis to the period beginning in August 2016, 60 events remain eligible for analysis. Of those, 59 retain both analytical windows and therefore enter the final event-level dataset. The final sample contains 24 fiscal events, 16 monetary policy/Copom events, 10 external events, and 9 political events.

Market data come from daily Bloomberg series for the DI curve and selected controls. The baseline analysis focuses on the DI 252d and DI 504d maturities and on the 21d–504d slope. Additional controls capture exchange-rate changes, oil prices, the VIX, Brazil five-year CDS, and U.S. Treasury yields. Focus expectations are merged using the last available observation prior to the relevant window, thereby avoiding look-ahead bias.

The event calendar links each shock date to the previous Copom statement and to the first subsequent Copom statement. The crucial implementation detail in the updated version is that if the shock and the Copom statement occur on the same calendar date, the statement is treated as posterior communication and its market response is measured on the next trading day. This rule aligns the empirical design with the institutional timing of Copom communication and prevents same-day Copom events from being dropped mechanically.

Two analytical windows are then defined. The first window, pre_shock_to_shock, runs from the last market day before the shock to the first market day on or after the shock. The second window, shock_to_statement, runs from that effective shock date to the first market day after the subsequent Copom statement. In the updated sample, the first window has the expected event-study profile, with a median length of one day. The second window is more heterogeneous, with a median length of 17 days overall, but a median of just one day for the monetary policy/Copom subgroup.

*Table 1 about here.*

*Table 1. Sample composition of complete events.*

| Shock type | Events | Share (%) |
| --- | --- | --- |
| Fiscal | 24 | 40.7 |
| Monetary Policy/Copom | 16 | 27.1 |
| External | 10 | 16.9 |
| Political | 9 | 15.3 |

## 4. Textual Feature Extraction

The textual component of the framework is deliberately structured. Each Copom statement is split into sentences and each sentence is classified into one of four categories: hawk, dove, neutral, or out of scope. The sentence-level classifications are then aggregated into a document-level tone measure. This procedure produces an interpretable score rather than a black-box text embedding.

A second layer of textual extraction is used to capture forward-guidance and uncertainty dimensions. Four variables are built for each statement: guidance direction, guidance explicitness, uncertainty level, and change in uncertainty. Guidance direction and explicitness are multiplied to obtain a composite guidance score. This design is useful for an applied paper because it maps the statement into variables that are economically legible and can be discussed alongside curve movements.

Across the sample, the text tone series becomes visibly more hawkish in the inflationary and fiscally uncertain episodes after 2021, remains positive through much of 2024–2025, and stays moderately positive in the most recent observations. This time series does not, by itself, prove market impact; however, it provides a transparent summary of the communication environment in which the event-study windows are embedded.

*Figure 2: time series of the hawk–dove score for Copom statements.*

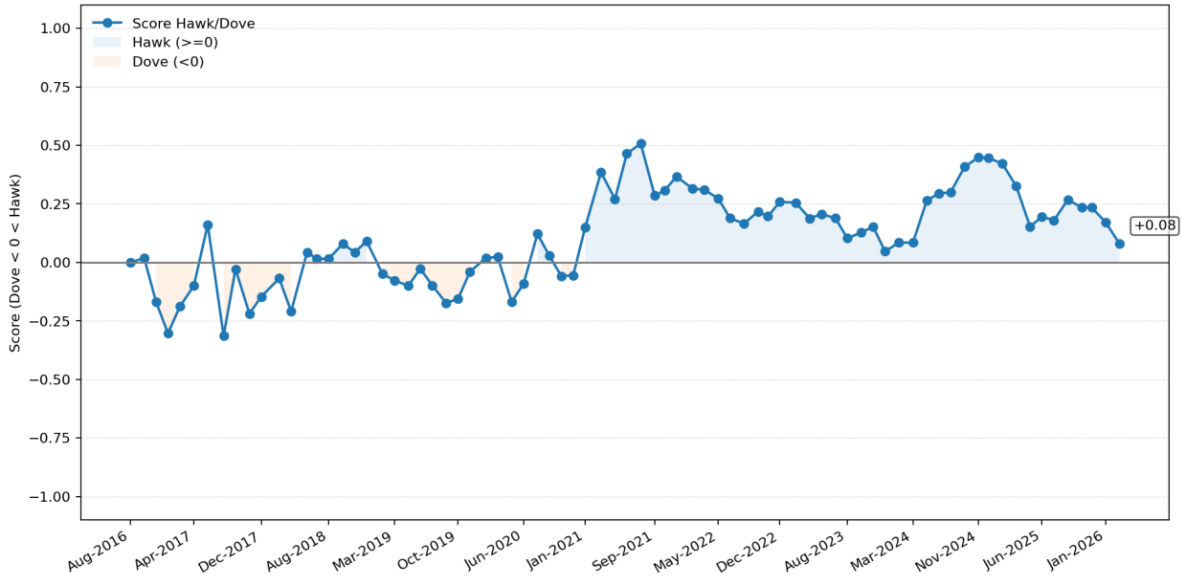

## 5. Empirical Strategy

The unit of analysis is the event. After the long event-window panel is built, the data are collapsed into an event-level dataset with one row per event. Variables from the first window receive the suffix _shock, while variables from the second window receive the suffix _statement. The dependent variables are therefore the repricing measures observed in the communication window.

For each target maturity m, the generic regression can be written as:

$$StatementRepricing_{i,m} = \alpha + \beta \cdot InitialShock_{i,m} + \Gamma'MarketControls_i + \Theta'Text_i + \Pi'Expectations_i + \varepsilon_i$$

Here, $InitialShock_{i,m}$ measures the repricing in the first window for the same maturity or slope, $MarketControls_i$ capture financial conditions observed in the shock window, $Text_i$ contains the structured communication measures, and $Expectations_i$ contains pre-event Focus variables. The purpose of the regression is not to identify a causal communication shock in the strict sense, but to assess whether the statement-window repricing can be empirically decomposed using information from the initial shock and the content of the subsequent Copom statement.

Baseline models are estimated by OLS with HC3 robust standard errors. Ridge and Lasso specifications, together with leave-one-out Ridge metrics, are used as complementary exercises to assess whether the framework has any meaningful out-of-sample stability. Because the sample remains modest, the number of regressors is kept intentionally small and targeted.

## 6. Results

### 6.1. Window behavior and event-level relationships

The first descriptive result is that the two windows capture different objects. The *pre_shock_to_shock* window is short and concentrated, consistent with a standard event-study measure of the immediate market reaction. By contrast, the *shock_to_statement* window is broader and more heterogeneous, which is consistent with the idea that it captures subsequent repricing in a communication window rather than a pure intraday statement surprise.

A useful empirical check is the relation between the initial shock and the subsequent statement-window move. For the DI 252d maturity, the simple correlation between the two measures is -0.03, indicating that the second-stage repricing is not mechanically proportional to the first-stage move. This feature supports the logic of modeling the second window explicitly rather than treating it as a trivial continuation of the initial shock.

*Figure 3: distributions of DI repricing by analytical window.*

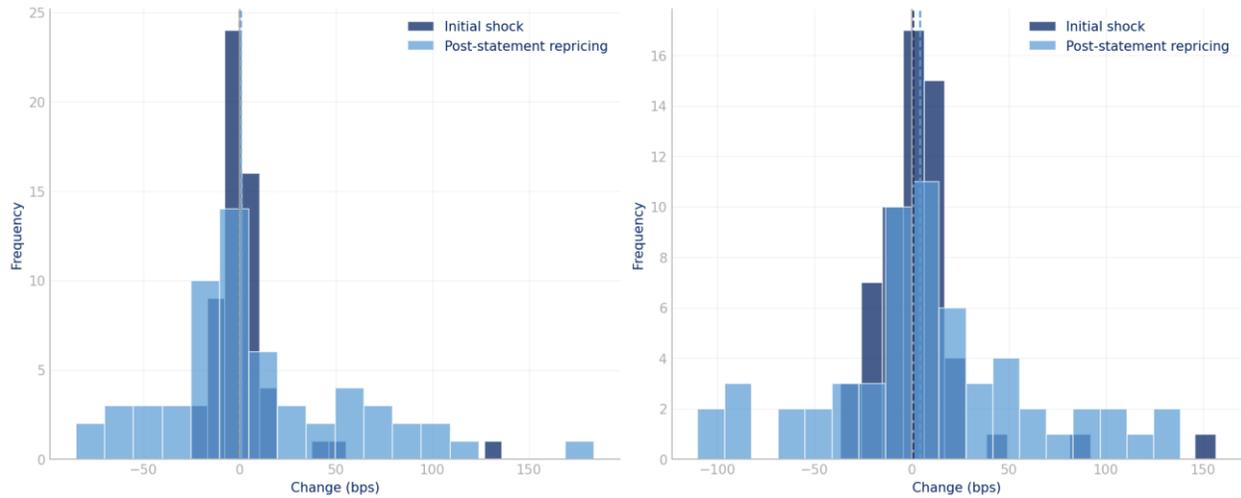

*Figure 4: scatter plot of the initial DI 252d shock versus statement-window DI 252d repricing.*

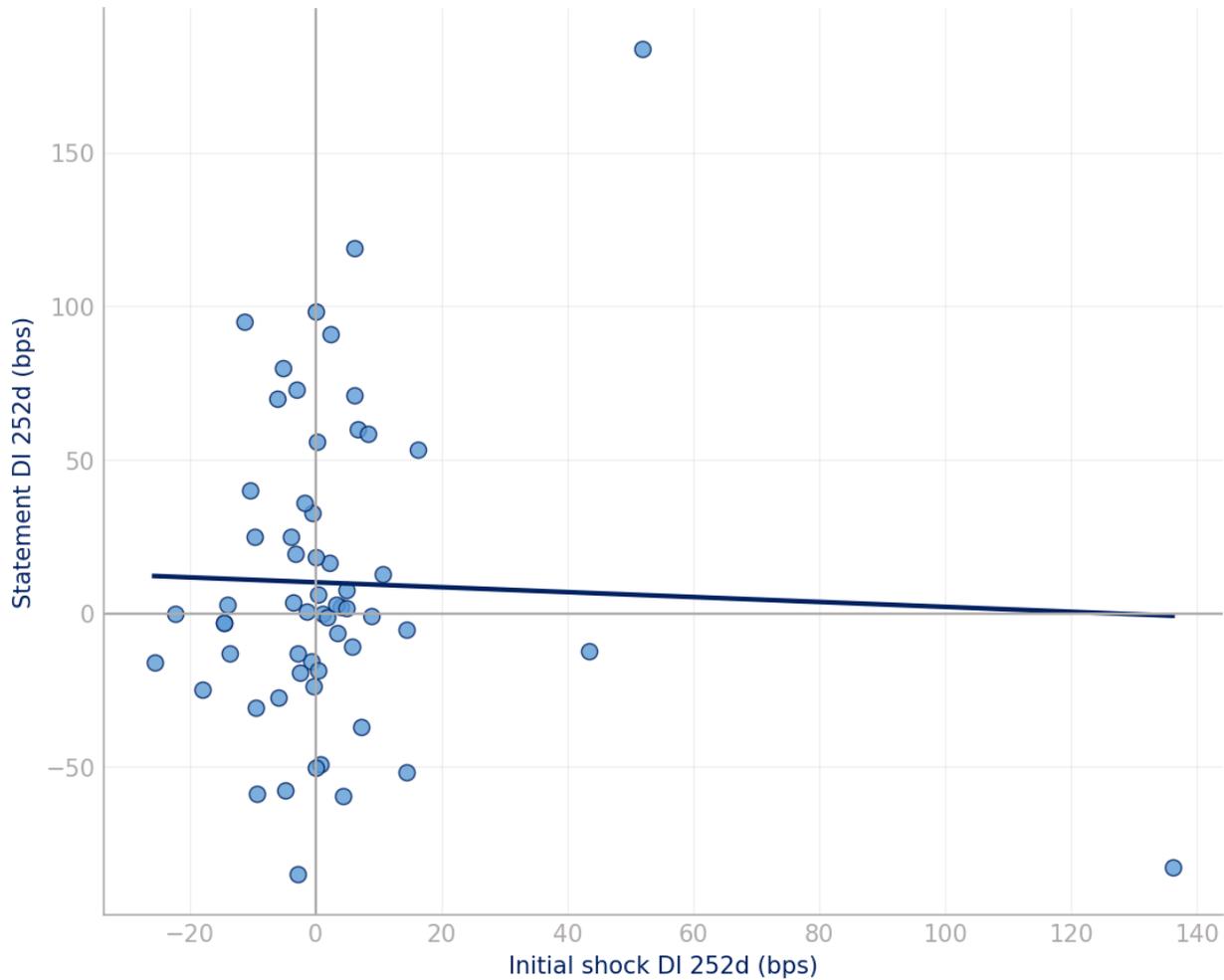

### 6.2. Baseline regressions

The baseline OLS-HC3 results are strongest for the DI 252d equation. The baseline specification reaches an in-sample $R^2$ of 0.432, with RMSE of 37.8 basis points and sign accuracy of 74.6%. The same equation also yields the clearest role for pre-event expectations: selic_year_pre enters with a negative and statistically significant coefficient ($p = 0.006$), which is consistent with the idea that richer pre-existing monetary tightening expectations leave less room for additional front-end repricing after the statement.

The same baseline exercise is materially weaker for the DI 504d equation ($R^2 = 0.055$) and for the slope equation ($R^2 = 0.099$). This maturity pattern suggests that the present framework is more informative for front-end or intermediate repricing than for longer maturities or slope adjustments, at least in the current sample and model space.

The coefficients of the textual variables are directionally plausible in the front-end equation. In the DI 252d baseline, a more hawkish tone is associated with higher statement-window repricing (coefficient = 44.6), and the guidance score also enters positively (coefficient = 8.9). However, the corresponding p-values (0.475 and 0.479) indicate that these effects are not estimated precisely enough to support strong causal claims. The correct interpretation is therefore that the textual variables are economically informative, but not uniformly robust.

*Table 2 about here.*

*Table 2. Main model performance metrics.*

| Specification | Estimator | N | R^2 | RMSE | Sign accuracy (%) |
|---|---|---|---|---|---|
| DI 252d + fiscal interaction | OLS-HC3 | 59 | 0.352 | 40.3 | 71.2 |
| DI 252d + fiscal interaction | Ridge LOO | 59 | 0.051 | 48.8 | 64.4 |
| DI 252d baseline | OLS-HC3 | 59 | 0.432 | 37.8 | 74.6 |
| DI 252d baseline | Ridge LOO | 59 | 0.085 | 47.9 | 67.8 |
| DI 504d baseline | OLS-HC3 | 59 | 0.055 | 53.2 | 61.0 |
| DI 504d baseline | Ridge LOO | 59 | -0.045 | 56.0 | 57.6 |
| Slope 21-504 baseline | OLS-HC3 | 59 | 0.099 | 54.3 | 52.5 |
| Slope 21-504 baseline | Ridge LOO | 59 | -0.052 | 58.7 | 61.0 |

## 6.3. Heterogeneity and robustness

The heterogeneity exercises are informative but should be interpreted carefully. In the current implementation, only the fiscal subgroup is large enough to sustain a separate regression exercise under the model's minimum-sample rules. Within that subgroup, the DI 252d specifications perform relatively well, but the evidence is not broad enough to claim stable heterogeneity across all shock types.

The robustness results reinforce the applied nature of the framework. For the DI 252d baseline, the leave-one-out Ridge R² is only 0.085, far below the in-sample OLS fit, and performance is weaker still for the DI 504d and slope targets. This divergence indicates that the framework is better understood as an explanatory decomposition tool than as a fully robust forecasting model.

*Figure 5: post-statement DI 252d repricing by shock type.*

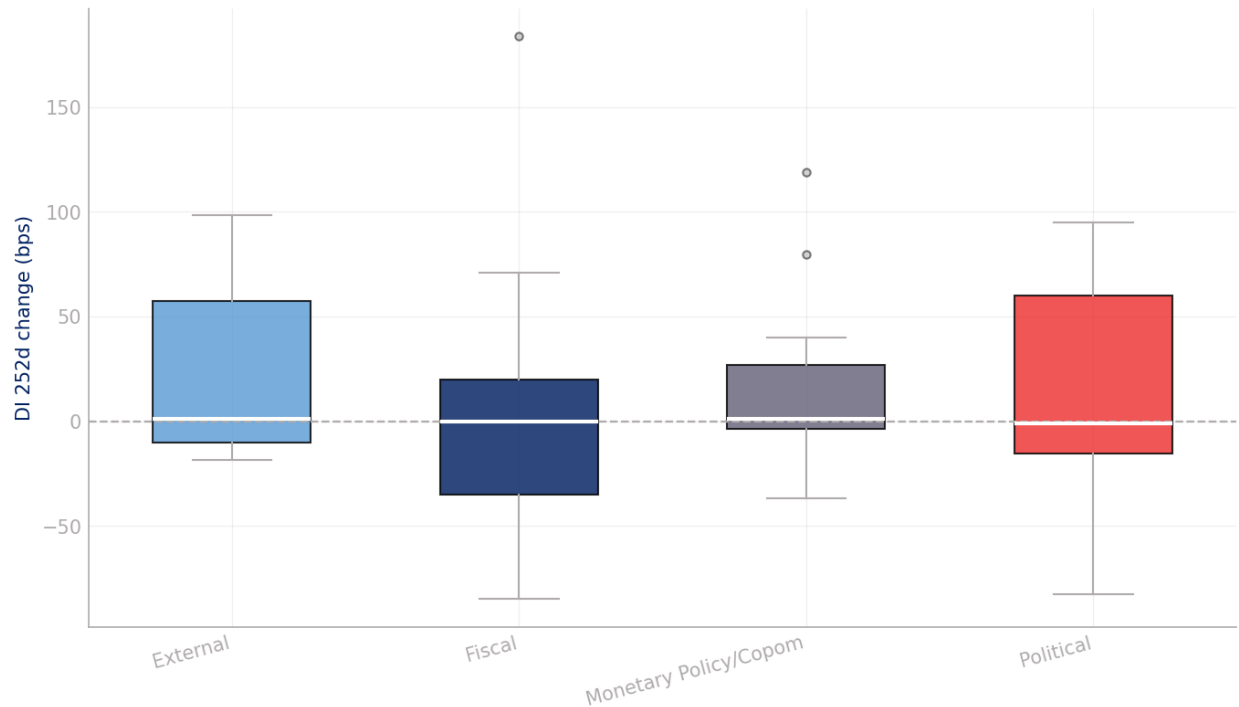

## 7. Discussion and Limitations

The paper's main strength is transparency. Each step of the framework is economically interpretable: the event calendar is explicit, the windows are observable, the Focus variables are merged without look-ahead bias, and the textual features have direct policy meaning. This makes the approach suitable for academic work with an applied orientation as well as for market-oriented policy analysis.

At the same time, several limitations should be acknowledged. First, the sample remains small. Second, the second window is daily rather than intraday, so it cannot fully isolate a pure communication shock from all intervening information. Third, the event categories are heterogeneous and some subgroups are too small for stable separate estimation. Fourth, the textual features are structured but still depend on a classification pipeline, which introduces measurement choices. Finally, the weaker cross-validated performance cautions against presenting the framework as a forecasting model.

These limitations improve rather than weaken the paper when stated clearly, because they align the empirical claims with what the design can genuinely support. The strongest claim is not that the model identifies the communication effect exactly, but that it offers a disciplined way to decompose yield-curve movements around Copom communication into an initial shock stage and a subsequent communication-associated repricing stage.

## 8. Conclusion

This paper proposes a two-step empirical framework for studying the Brazilian yield curve around Copom-related events. The framework separates the initial market repricing associated with the underlying shock from the subsequent repricing observed in the communication window that follows. By combining event-study logic, structured textual variables, market controls, and survey expectations, the paper offers a practical decomposition that is useful for applied monetary policy analysis.

The updated implementation materially strengthens the design by preserving same-day Copom events in the communication stage, which produces a complete event-level sample of 59 observations with balanced windows. The empirical results suggest that the framework is most informative at the front end of the curve, especially for the DI 252d maturity. Communication variables appear to contain economically meaningful information, but their effects are not sufficiently stable to justify strong causal or forecasting claims.

The most credible way to position the paper is therefore as a technical-applied contribution in academic form. It delivers a serious empirical framework, documents what the data can and cannot say, and opens a clear agenda for future work using higher-frequency windows, richer event stratification, and alternative text measurement strategies.

# Appendix: Figure Placement Roadmap and Previews

*Appendix Table A1. Recommended figure plan.*

| Figure | Placement in manuscript | Content | Use | Why it helps |
|---|---|---|---|---|
| Figure 1 | Section 1 / end of Introduction | New schematic timeline of the two-step event design | Main text | Conceptual anchor for the identification strategy. |
| Figure 2 | Section 4 / end of Textual Feature Extraction | Hawk–dove score history of Copom statements | Main text | Shows the evolution of communication tone across the sample. |
| Figure 3 | Section 6.1 | Distributions of DI 252d and DI 504d repricing by window | Main text | Illustrates that the two windows capture different empirical objects. |
| Figure 4 | Section 6.1 | Scatter of initial DI 252d shock vs statement-window DI 252d repricing | Main text | Shows weak mechanical relation between the two stages. |
| Figure 5 | Section 6.3 | Boxplot of post-statement DI 252d repricing by shock type | Main text | Useful heterogeneity visualization without strong parametric claims. |
| Figure A1 | Appendix | RMSE comparison across estimators | Appendix | Supports the claim that forecasting robustness is limited. |
| Figure A2 | Appendix | Text coefficients by shock type | Appendix | Diagnostic only; subgroup evidence is limited. |
| Optional Figure A3 | Appendix | Baseline coefficient forest plot (new) | Appendix | More academic alternative to the generic top-coefficients chart. |

*Figure A1: RMSE comparison across estimators.*

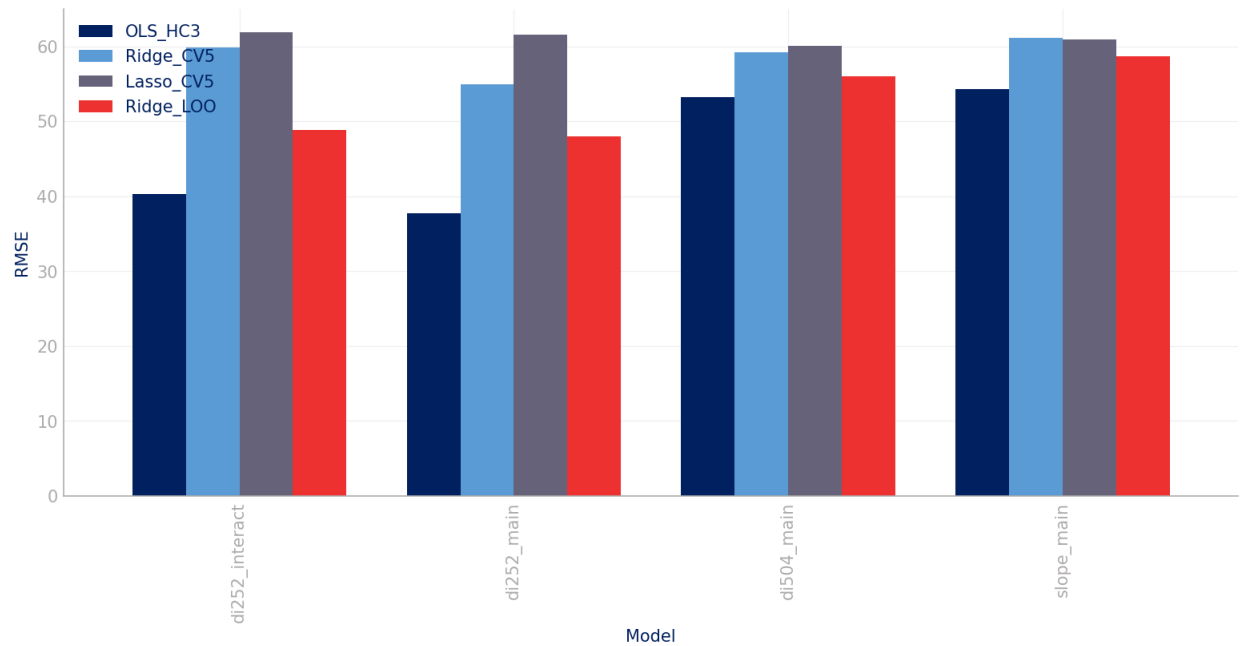

*Figure A2: textual coefficients by shock type.*

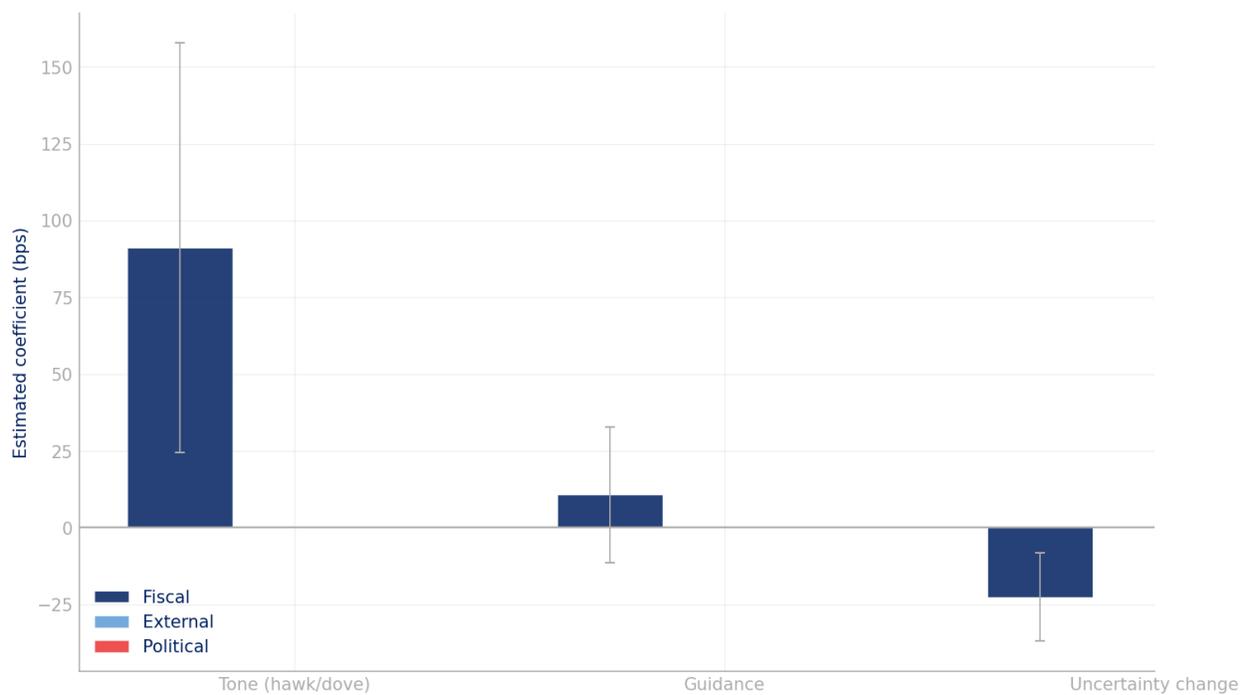

*Optional Figure A3. Baseline OLS-HC3 coefficients with 95% confidence intervals.*

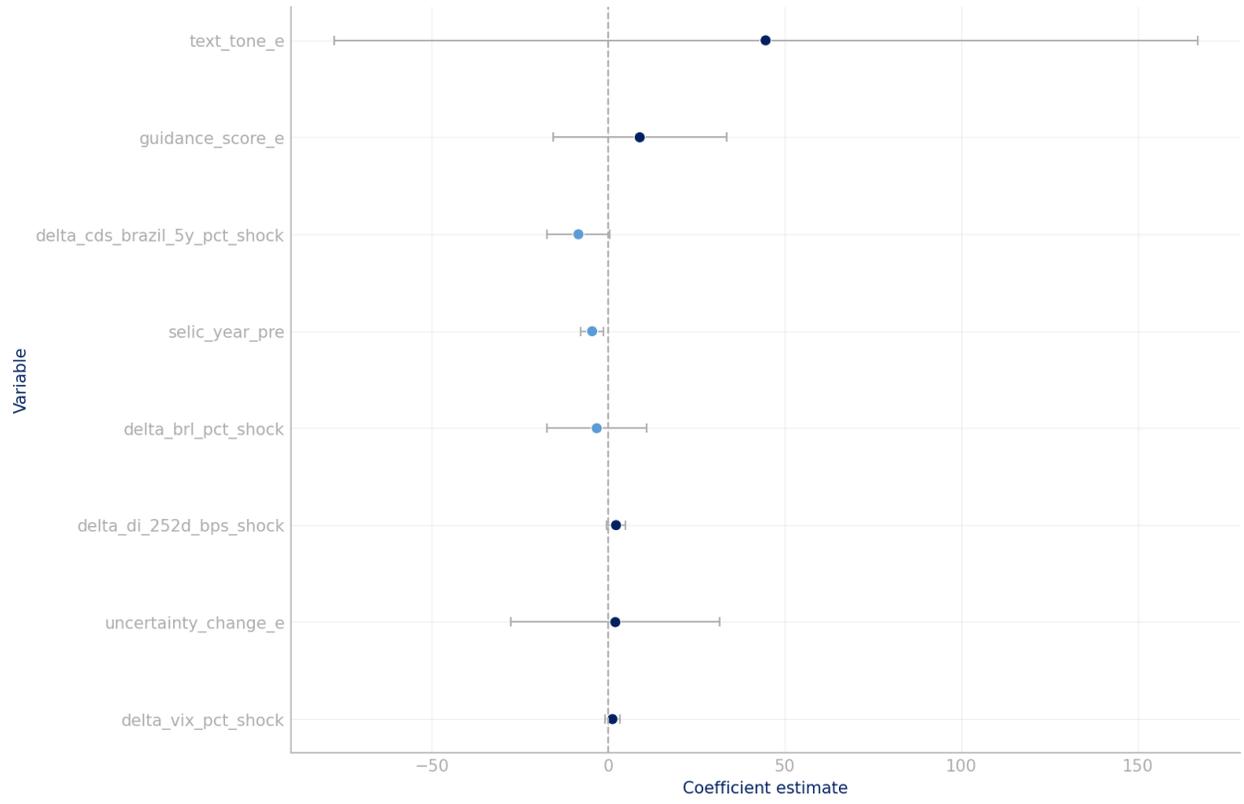